# IDEALI: intuitively localising connected devices in order to support autonomy

Réjane Dalcé[a], Antonio Serpa[b], Thierry Val[c], Adrien van den Bossche[c], Frédéric Vella[b], Nadine Vigouroux[b]

[a] *Institut de Recherche en Informatique de Toulouse, Institut National Universitaire Champollion, France*
[b] *Institut de Recherche en Informatique de Toulouse, Université Toulouse 3, France*
[c] *Institut de Recherche en Informatique de Toulouse, Université Toulouse 2 Jean Jaurès, France*

**Abstract**
The ability to localise a smart device is very useful to visually or cognitively impaired people. Localisation-capable technologies are becoming more readily available as off-the-shelf components. In this paper, we highlight the need for such a service in the field of health and autonomy, especially for disabled people. We introduce a model for Semantic Position Description (SPD) (e.g. "The pill organiser in on the kitchen table") as well as various algorithms that transform raw distance estimations to SPD related to proximity, alignment and room identification. Two of these algorithms are evaluated using the LocURa4IoT testbed. The results are compared to the output of a pre-experiment involving ten human participants in the Maison Intelligente de Blagnac. The two studies indicate that both approaches converge up to 90% of the time.
.

## 1. Introduction

While finding out the location of a given object has been a constant issue in human life, technological solutions that provide an answer in an indoor environment are now emerging. In fact, until recently, location determination implied the use of Global Positioning System (GPS) or, more precisely, Global Navigation Satellite System (GNSS) technologies. Todays off-the-shelf components provide a service that is complementary to the satellite-based approach: low to moderate cost indoor localisation with centimetre-level precision becomes feasible and opens the way to more advanced studies. The work described in this article targets the application of modern localisation techniques to the field of health and personal autonomy and discusses the technological needs and possibilities.

We will start by introducing the needs for object and people localisation in the field of assisted living. Afetr a brief overview of the existing localisation and geolocalisation technologies, we introduce the concept of Semantic Position Description (SPD) of an object as a response to the localisation challenge. We report first results obtained with the LocURa4IoT platform present on the Blagnac Smart House. We then describe the pre-experimental protocol implemented to identify how users describe the localisation of objects in relation to reference objects. Then we confront the results of the location algorithm to the users' descriptions. Finally, we discuss perspectives such as the analysis of location adverbs and the proposal of new localisation algorithms.

## 2. Needs and state of the art of guidance, identification and location of objects for people with disabilities

### 2.1. Locating objects: a real need

The process of identifying and locating objects can be difficult and stressful for people with disabilities. At the same time, it is a key element in maintaining their autonomy. In this section, we will demonstrate this for two disability situations: visual impairment and cognitive impairment.

In the field of visual impairment, the survey of 54 blind people by Dramas et al [4] has shown that the ability to locate objects for blind people corresponds to a real need and contributes to increasing their autonomy. However, this localisation requires the blind person to acquire a representation of space. These same authors postulate that continuous localisation, based on local cues, allows one to have a representation of space in order to follow, reach and grasp the desired object.

In the field of cognitive impairment, [5] and [6] report that misplacing objects is a symptom of Alzheimer's disease. In addition, the non-profit, non-governmental organisation Alzheimer Europe [7] emphasizes the effect of object loss on the social relationship between the patient and his or her caregivers.

Visually impaired people need guidance because of the lack of perception of objects in space. Cognitively impaired people also need guidance because of concentration/memorisation /understanding disorders.

Based on these observations, Figure 1 represents the need for an object localisation function which must be part of an assistive system. This will rely on appropriate interaction modalities and technologies in order to provide adequate spatial information regarding the object or target's location. The object location technologies will inform the entities in charge of user interaction about the position of the object. This will be used by navigation systems for the visually impaired, for example, or by cognitive assistance systems to facilitate the achievement of ADLs [1] or IADLs [2] for people with cognitive degeneration.

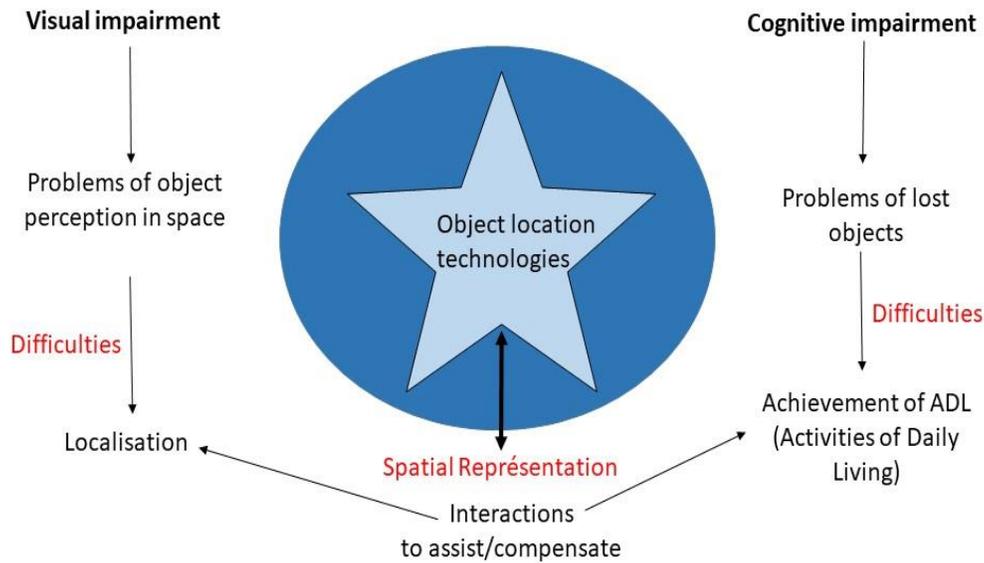

Figure 1.The need for object location in the field of autonomy.

### 2.2. The spatial reference

Object localisation requires the definition of a spatial reference. Shelton and Mac Namara [8] define "a spatial reference frame" as "a system of relationships consisting of the location of objects, reference points, and the spatial relationships that may exist between them". There are two main types of spatial reference frames: egocentric and allocentric [9], [10], [11], [12].

In the egocentric frame of reference, the location of the object is encoded according to the position of the observer, whether static or moving. The point of view depends on the position and orientation of the subject respectively defined as the origin and the reference axis of any distance or angle measurement. The angle formed by the subject's orientation (head or body reference frame) and the direction of an element of the environment is called "bearing" and is expressed with specific adverbs (right, left, in front, behind, etc.), in degrees (90 degrees to the left, for example), or even, in a particularly intuitive way, especially for visually impaired people, by using the metaphor of the clock [13]. On the other hand, the allocentric type of reference frame is independent of the subject's point of view. The information is expressed as a function of the position of one or more fixed objects in the environment and therefore does not vary with the position of the observer (e.g. "the town hall is opposite the bakery"). In this frame of reference, the encoding is done according to several distant landmarks that provide a frame of reference from which the object is located in terms of distances and directions.

### 2.3. A new approach for indoor-localisation based on semantics

Semantic-based indoor localisation : a new approach

When it comes to navigation, localisation and identification of objects for visually impaired and/or cognitively impaired people, a number of technologies have already been explored (Tab. 1). This table also lists the situations of use. These studies often address indoor/outdoor navigation applications for the visually impaired using voice, vibro-tactile feedback or augmented reality.

Our approach is slightly different as we aim to provide more intuitive feedback to the user. The proposed system:

- Produces a Semantic Position Description (SPD) using the estimated distances between an object of interest and known reference points. An SPD is a string describing an aspect of the position of an object. For example, "your keys are close to the coffee pot on the left of the fridge" is an SPD describing the position of the keys based on the coffee pot and the fridge,
- Aims at modelling the state of each object at any point in time using the various sensors deployed in the environment. This aspect will allow our proposed system to take the context into account in its interaction processes. These processes will then be more intelligible to the end user, thus more efficient, which is a key feature when considering their application to the field of autonomy. The information can therefore be tailored to the user's profile and context of use. Specifically, we hypothesize that, in some situations, information such as "object A is between object B and object C" is more relevant than three sets of coordinates [$\{x_A, y_A, z_A\}, \{x_B, y_B, z_B\}, \{x_C, y_C, z_C\}$] or angles and distances.

## 3. Localisation and geolocalisation technologies

Using the existing litterature, the previous section highlighted the need for identification, guidance and localisation systems for people with disabilities. This section gives a quick overview of the available technologies that may meet these expectations.

### 3.1. General Concepts of localisation and positionning technologies

#### 3.1.1. Range, referential and spatial reference

A key aspect when designing a localisation system is the target coverage. In the literature, two main classes can be identified: indoor and outdoor localisation systems. The distinction between the two is based on the following criteria:
- Deployment location: depending on whether the devices are placed inside or outside of buildings, specific classes of network technologies (WWAN, satellite, WLAN, WPAN...) can be considered.
- Coordinate system: the reference frame used to encode the position of the mobile can be universal and global, or local, expressed in spherical, cartesian/orthonormal coordinates, etc. For outdoor use and for the sake of universality over the entire globe, the reference frame will be spherical. For an indoor localisation, a local coordinate system, generally orthonormal - although this may depend on the shape of the buildings - will frequently be used. Rules for changing the reference frame make it possible to switch from one to the other in order to imagine "seamless" localisation services.

A key aspect when designing a localisation system is the target coverage. In the literature, two main classes can be identified: indoor and outdoor localisation systems. The distinction between the two is based on the following criteria:

Tab. 1. Technologies and uses for object and person localization.

| Approach Population | Technologies used | | Examples of uses | References |
|---|---|---|---|---|
| | Navigation | Localisation and/or identification of objects/persons | | |
| Cognitive impairments | | Humanoid robot Optical acquisition system | Object localisation (voice feedback) | [15] |
| | | Spatialised sound RFID tracking | Object localisation | [16] |
| Visual impairments | | RF Tags | Object localisation Indoor/outdoor navigation | [17] |
| | | Ultrasound/infrared | Indoor/outdoor navigation Detection of risk sources Object/Pattern Recognition | [18] |
| | | Scene analysis by computer vision | Automatic text analysis Object location, Face identification, Bar code reading, QR code etc. Detection of risk sources | [14][19] |
| | | Prosthetic vision | Object/pattern/person location Object/pattern/person recognition | [20] |
| | | Augmented reality video and/or audio | Indoor/outdoor navigation Object location Risk/obstacle detection | [21] |
| | | Vibro-tactile | Target localisation Indoor/outdoor navigation Risk/obstacle detection | [22][23] |

*3.1.2. References and anchors*

To enable localisation, the system generally needs reference points with known positions: these points are called "reference objects" or "anchors". These anchors should also be easy to recognize in order for information mentioning them to make sense. The quality of their semantic identification and position is critical to the accuracy and reliability of the entire location system. As stated in the previous chapter, when the reference frame is user-centric, the term ego-centric will be used. In this case, the position and/or the distance to the object of interest is defined with respect to oneself (the user). If the origin of the frame of reference is independent of the pair (user; mobile), the localisation will be termed allo-centric and will be defined in the coordinates of the frame of reference.

*3.1.3. Architecture*

The presence of an infrastructure is also a differentiation criterion for localisation systems. The GNSS is an example of an infrastructure based system: the satellites that are used by mobile devices to compute their position orbit the earth whether or not the mobiles are present or active. On the other hand, the StopCOVID app used in France in 2020 for relative proximity detection, showcases an infrastructure-less localisation system: an information of close proximity, without using position expressed in an absolute referential, is generated in an opportunistic manner based on simple Bluetooth signal strength. The proximity detection service therefore does not rely on existing communication infrastructures such as 3G/4G/5G or GPS.

*3.2. Overview of common localisation technologies and methods*

In this section, we briefly present the different approaches (principles, observable quantities, technologies) used to achieve connected devices localisation. We present historical solutions as well as recent advances in the domain.

**Global Navigation Service** (GNSS): this universal service achieves a global range thanks to a satellite-based infrastructure covering the entire globe. The performance decreases indoors and in urban canyons because of the Non-Line-Of-Sight (NLOS, presence of obstacles) between the satellites and the mobile: in severe cases, the service may become unresponsive. An accuracy of a few meters is achievable in the best-case scenario, in a public use.

**Vision:** the use of cameras, combined with image processing, allows the recognition and location of objects and people. To ensure operation at night, it is possible to use infrared (IR) cameras. This is the case of active 3D IR cameras such as the well-known Kinect, which generates a 3D image of the observed scene [24]. This localisation system may lack accuracy in some situations because it is sensitive to changes in lighting conditions, salience and/or apparent size of the objects to be recognized.

**Inertial sensors:** accelerometers, gyrometers and 3D magnetometers are often combined in an IMU (Inertial Measurement Unit). The first two sensors provide measurements that are linked to relative movements. It is therefore necessary to couple them to an initial position/orientation, which can be corrected by using a magnetometer, given that the environment allows such measurements. However, even with potentially heavy processing, the accuracy is limited and deteriorates over time unless another system is used to mitigate the impact of error accumulation due to integrals [25].

**Laser/Infrared/Ultra-Sound:** solutions coming from the world of robotics are interesting, such as the use of Lidar (remote sensing by laser), infrared or ultrasonic signals [27]. The first solution allows to locate objects, not necessarily connected, in 2D or 3D. It is also possible to use simple PIR (Passive InfraRed) sensors such as those used by domotic alarms to detect the temperature variation in a room: they may also serve as an indicator of the presence of a moving person for example [26]. Unfortunately, the complexity of such solutions grows quickly when the system must support multi-user scenarios: the number of PIR sensors is significant and another technology is still required to identify each person [26].

**RFID/NFC:** RFID (Radio Frequency Identification) tags - passive or active - or more recently NFC (Near-Field Communication) tags have been exploited for many years to locate objects, mainly in indoor environments. This system requires the deployment of numerous tags on objects or in the environment that react to signals generated by a specific reader: this results in a rough position estimate, based on proximity. The short range of these radio-frequency technologies (especially for NFC with a few cm range) requires a very dense mesh, which translates into a high number of connected devices. These constraints make this technology complex to implement and not very accurate. RFIDs (especially active ones) offer ranges of several meters, but the localization process, based mainly on the received power measurement, is not very accurate.

Nowadays, mainstream localization technologies are mainly based on the following measurement methods:

**Received Signal Strength Indication (RSSI):** this well-known parameter is based on the evaluation of the power of the incoming radio signal. Its success mainly comes from the fact that it is available on most radio receivers. By using a standard path-loss model and knowing the transmission power, the distance can be estimated using the RSSI. The downside is the reliability of the distance

estimate: radio signals are influenced by the propagation environment, therefore the presence of obstacles, whether static (wall, furniture...) or mobile (human...) results in significant errors due to NLOS [28], [29].

**Time-of-Flight (ToF):** measuring the time taken by a radio message between the transmission and reception antenna is the most accurate way to evaluate a distance between transmitter and receiver. For indoor localisation, UWB (Ultra-Wide Band) allows centimetre-level accuracy, especially in LOS situations. The ToF offers a ranging or distance estimate that can then be fed to localisation algorithms such as trilateration [30].

**Angle of Arrival (AoA):** this goniometry system dates back to the Second World War where it was used to evaluate the angle of arrival of radio signals outdoors. It is also used by mobile operators to triangulate GSM (Global System for Mobile) signals, e.g., during the search for persons after an avalanche [31]. Unfortunately, the antenna arrays or rotating antenna required by traditional AoA are too cumbersome for indoor applications. The latest developments in Bluetooth from v5.1 aim to provide angle of departure (AoD) control and AoA evaluation by providing access to radio IQ parameters [32]. Qorvo, the world leader in UWB, is also working on AoA with their latest UWB radio tranceivers [33].

At the time of writing of this document, Qorvo's UWB transceivers represent the state of the art regarding the ability to measure both distance and angle of arrival. This allows a single well-positioned and oriented DW3000 UWB receiver to know the absolute location of a UWB radio transmitter, typically in a room. The accuracy of the angle measurement is around 10 degrees.
.
Tab. 2 summarizes the presented technologies and Tab. 3 covers the informations that can be exploited for ranging/localisation.

It is interesting to note that many research studies use several techniques together. The accuracy is then improved, but this is at the cost of a greater complexity, which may come from longer computation times, a greater processing power, a larger memory footprint, a higher energy consumption and/or a higher pricetag.

In our study, we selected UWB for its achievable accuracy on the one hand, but also for the numerous possibilities offered (ToF for distance measurement and AoA for direction detection).

In LOS situations, UWB offers an accuracy of a few centimeters, without heavy processing: the raw, instantaneous data often present sufficient accuracy for localisation. The radio coverage reaches up to 100m, and the time measurements are less disturbed by signal attenuation compared to RSSI. The interested reader may consult [38] for more details.

These features will definitely be useful when designing both allocentric and ego-centric approaches.

Tab. 2. Main localisation technologies

| Name | Features | Environment | Infrastructure | Precision | information used |
|---|---|---|---|---|---|
| GNSS | universal/global, spherical marker (latitude, longitude, altitude) | *outdoor* | satellites | metric | ToF |
| Vision | Cameras | *indoor* | cameras | low | pixel analysis |
| Inertial | Accelerometers 3D gyrometers | *indoor* | no | relative | motion |
| Laser | Lidar, lasers | *outdoor* | cameras | medium | ToF |
| RFID/NFC | Proximity detection | *indoor* | reader | low | RSSI |

Tab. 3. Main localisation informations

| Name | Features | Environment | Infrastructure | Precision |
|---|---|---|---|---|
| motion | measurement of linear and angular displacements... | indoor / outdoor | no | low |
| image | image analysis | indoor / outdoor | camaras | low |
| RSSI | Distance measurement | *indoor* | beacons | metric |
| ToF | Distance measurement between 2 nodes | *indoor* | beacons | cm |
| AoA | Detection of reception angle | *indoor* | beacons | 10 degrees (UWB) |

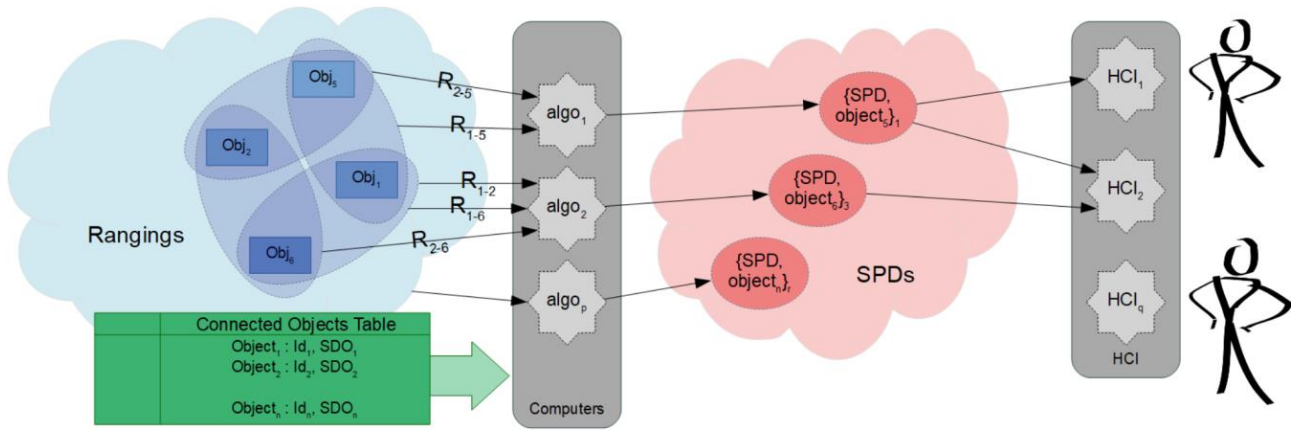

Figure 2. System architecture.

*3.3. Output of the localisation system and its use*

Localisation systems produce a wide variety of outputs: GNSS coordinates, distances, positions in the local orthonormal reference frame, etc. This information is often indigestible to the end user: a complementary technology like a map or an interpreter is required to make sense of the data. The remainder of this article is focused on this topic.

In addition, many studies focus on obtaining the location data but not on the protection of said data, especially from a confidentiality point of view. Knowing the objects' location may allow an attacker to determine the person's location. In this respect, as in all work involving connected objects, the security question is legitimate and the user must be provided transparent information about the storage and use of the data produced, and a right of withdrawal.

## 4. The IDEALI project

The aim of the IDEALI projet "Identification Assisted by Localisation for Smart Home" (in French "IDEntification Assistée par la Localisation dans l'habitat Intelligent") is to facilitate the use of the localisation information of a connected object used in everyday life by making it more palatable to the end user.

First, a localisation system based on UWB ToF has been proposed, designed and implemented: this system takes the form of a testbed [35] and added protocols and algorithms as illustrated in [37]. Its output has then been adapted to a user with a disability such as those described above (visual impairment or cognitive impairment) to help him find a lost object. In a second phase, IDEALI will make it possible to conceptualize interactive support systems that are more intuitive and adapted to the capabilities and activities of users. The first phase of the IDEALI project addresses the issue of Semantic Position Description (SPD), i.e., a description of the localisation of the connected object in a natural expression: the objective is to allow better assimilation and understanding of location information by the user. In AAL (Active Assisted Living) situations, this SPD will be much easier to interpret by a user than a {x,y,z} coordinate triplet.

The proposed IDEALI architecture, described in chapter V, was designed to produce an SPD from measurements processed by a set of algorithms. These algorithms rely on two input data:
1. Real-time ranging data (distances) between connected objects in the environment, obtained by ToF ranging protocols over UWB,
2. Semantic Object Description (SOD), i.e., semantic data associated with the connected objects involved in the ranging process.

We assume that each object known to the system is associated with an SOD, which is stored in a database. As a first approach, we considered that any connected object can initiate a ranging process or distance measurement with any other connected object in the environment. These ranging results are then made available to the algorithms via an MQTT bus [35] (Figure 2). For each connected object in the environment, the algorithms produce an SPD fragment based on both the distance measurements and the SOD of the objects involved in the measurement. This behavior aims at mimicking the use of several levels of position description that is done naturally by humans: for example, a person could describe his/her smartphone location as "It should be in the livingroom, on the table", indicating the room then a specific area of the room. Generating this position description requires a mechanism to select suitable reference points [34] as well as a sorting algorithm that will format the final data before presenting it to the end user. The proposed algorithms so far are independent from the caracteristics of the targeted population. In the future, we will co-design the use cases with the end users. In this paper, we focus on the SPD generation, therefore, the neighbour (reference point) selection and output formatting are out of scope.

## 5. Algorithm's description and evaluation

This section describes the algorithms implemented in the IDEALI project. The test results using the LocURa4IoT testbed are then presented.

### 5.1. Proposed Micro-algorithms

Several micro-algorithms have been proposed in the IDEALI project. In this paper, we present 3 of them: "Room Determination", "Proximity Estimator" and "Alignment Estimator" It should be noted here that because of the absence of user localisation and orientation information, the system can only provide allo-centered type information.

#### 5.1.1. Room Determination

The Room Determination algorithm is based on an observation of the closest wireless neighbours, i.e., connected objects in the wireless range of the target. The room where each neighbour is located can be found in the SOD. Therefore, the algorithm can use a majority vote - like mechanism to assign a room ID to the target using the closest neighbours.

#### 5.1.2. Proximity Estimator

The Proximity Estimator algorithm considers 1) the position of fixed and easily recognizable objects $A_j$, and 2) objects to be located $X_i$, potentially mobile in the environment. The distance measurement (ranging) between $X_i$ and $A_j$ provides proximity information. The Proximity Estimator compares the estimated distance to fixed thresholds in order to build the SPD: depending on the distance, $X_i$ and $A_j$ will be considered "very close" (VC), "near" (N) or "in the vincinity" (V) of each other. Distances greater than the upper limit of the "in the vincinity" case cause the algorithm to return NULL. Tab. 4 indicates the implemented mapping.

Tab. 4. Ranging thresholds used in the PE algorithm.

| Ranging (m) | Proximity |
|---|---|
| R < 0,3 | « Very close » |
| 0,3 < R < 0,6 | « Near » |
| 0,6 < R < 1,2 | « In the vicinity » |

When the Proximity Estimator is run with multiple $A_j$ for a single $X_i$, an option selecting the nearest $A_j$ can be activated. In this case, the Proximity Estimator returns only the closest proximity.

#### 5.1.3. Alignment Estimator

The Alignment Estimator algorithm indicates the existence of a rough alignment between three objects. Its output is therefore of the type "object $X_i$ is between $A_j$ and $A_k$". To solve the problem, the proposed approach consists in a bounding rhombus. Consider a triangle formed by $X_i$, $A_j$ and $A_k$. If the angles at $A_j$ and $A_k$ are below a given threshold, $X_i$ is between $A_j$ and $A_k$.

#### 5.1.4. SPD Combiner

For a given object, each algorithm i provides $SPD_i$, its fragment of SPD. These fragments must then be put together to produce a final SPD. In a first approach, we proposed a simple model where the final string is composed of each non-NULL $SPD_i$ separated by a comma. Using the three algorithms described before, an example of a combined output would be: "[object is] in the kitchen (1), near the sink (2), between the coffee maker and the sink (3)".

Information (1) is the output of the Room Determination algorithm, (2) the output of the Proximity Estimator, and (3) comes from the Alignment Estimator.

A first prototype has been implemented using site P2 of the LocURa4IoT platform [35] at the Maison Intelligente de Blagnac [36]. On this platform, 10 Ultra-Wide Band wireless devices are deployed in a local allo-centered reference frame. The first two algorithms described have been implemented, as well as the SPD combiner. A moving object, attached to a key ring, has been "misplaced" in multiple positions, and located by the system. The combined output was observed. Several examples are given in Tab. 5.

Tab. 5. Results achieved.

| Position | Final output of IDEALI |
|---|---|
| Few cm of television | *"In the livingroom, near the television"* |
| Few dozens cm of television | *"In the livingroom, in the vicinity of the television"* |
| On the couch, far from all others objects | *"In the livingroom"* |
| On the kitchen counter | *"In the kitchen, near the sink"* |

## 5.2. Experiment on testbed

In this section, we present the results of the Proximity Estimator and Alignment Estimator algorithms run on an automated testbed. The Room Determination Algorithm has not been experimented on the testbed, since the node's disposition on the testbed do not make this experiment possible.

The experiments were run using site P1 of the LocURa4IoT platform [35]. On this site, the testbed hosts around 60 radio devices which support both Ultra-Wide Band and Bluetooth Low Energy radio interfaces. During the experiments, the UWB radio was used. One of the devices is mounted on a rail and can move over a 7m span. At each end of the rail, a fixed node is also mounted: only one of those two nodes is represented on figure 3. In terms of relative positioning, the grid shown on Figure 3 has a 30cm resolution.

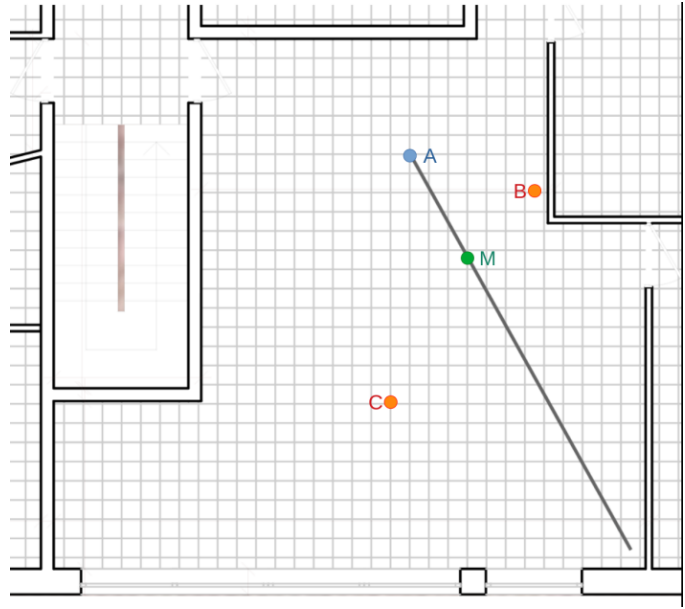

Figure 3. Fixed and mobile nodes used on the testbed.

A key component of the testbed is the MQTT bus. This bus facilitates data collection but also the deployment of software agents that analyse the results of the experiments. In the following, the micro-algorithms are connected to the bus and subscribe to the appropriate topics. Figure 4 illustrates the architecture used during the tests.

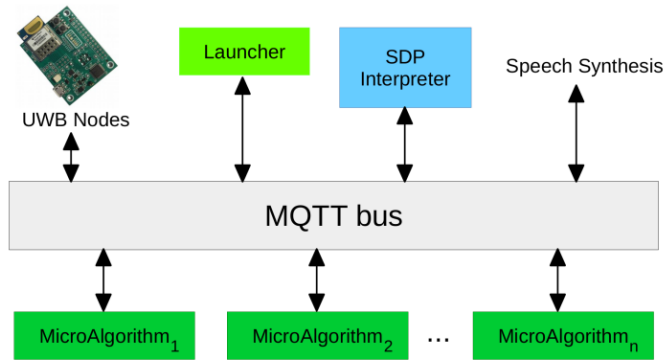

Figure 4. System architecture during the tests.

### 5.2.1. Proximity Estimator

The scenario used to evaluate the Proximity Estimator involves the mobile node M and one of the rail end nodes, A. During the experiment, M measures the distance from itself to A a thousand times and moves 25cm further. The ranging results are published on the MQTT bus and processed by the Proximity Estimator.

As previously stated, the Proximity Estimator compares the estimated distance to 3 threshold values in order to indicate whether the node (M) is "very close to", "near", or "in the vicinity of" the reference point (A). Therefore, the ranging accuracy is key to the algorithm performance. Figure 5 shows the ranging estimation with a focus on the 0-3m interval as M moves away from A. In the interval of interest, the value gets larger as the nodes get closer. Figure 5 makes it clear that a severe underestimation of the distance takes place: an object that is in the "near" interval (0.4m) will be reported by the system as being "very close to". Similarly, an object that is 0.75m away will be seen as "near" instead of "in the vicinity of" the reference point.

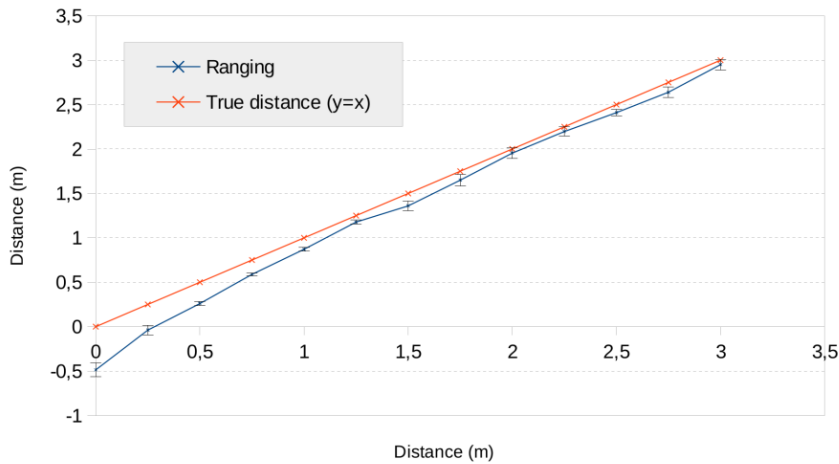

Figure 5. Ranging error.

### 5.2.2. Alignement Estimator

The Alignement Estimator algorithm tries to create a triangle between nodes M, B and C using the known and measured distances and the Al Kashi theorem: if the angles at vertices B and C are under 30°, nodes B, M and C are considered aligned.

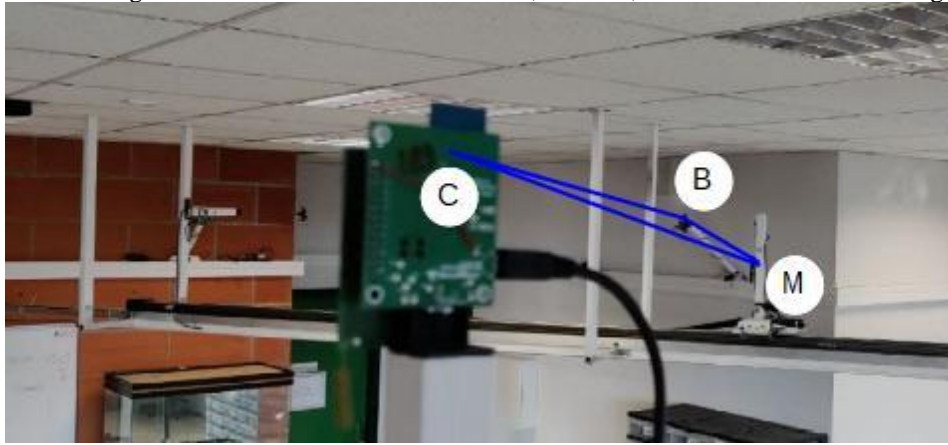

Figure 6. Alignment of the nodes on the testbed.

Figure 6 shows the nodes orientation. They are all mounted on a PVC structure 2.65m high. Node M is mounted on the rail and will move along this axis during the experiment.

In order to evaluate the algorithm, the mobile node of the testbed and two other nodes were selected. Nodes B and C are located each on one side of the rail. As with the previous scenario, M moves 25cm further between each ranging session. At some point, it passes between B and C.

The results were divided into three categories based on the computed angles:
- When the real angles are greater than the threshold, the algorithm correctly returns that the nodes are not aligned 69.53% of the time (out of 558 samples).
- When the real angles are below the threshold and the difference between the two is less than 10%, the success rate is 88,64% (out of 502 samples).
- When the real angles are very small and the nodes are visibly aligned, the original algorithm fails because of the aforementioned ranging error issues: the distance underestimation is such that the sum S of the estimated distances (B-M and C-M) is smaller than the distance between B and C (DBC) (Figure 7). The algorithm is thus unable to provide an SPD 93.63% of the time (out of 487 samples).

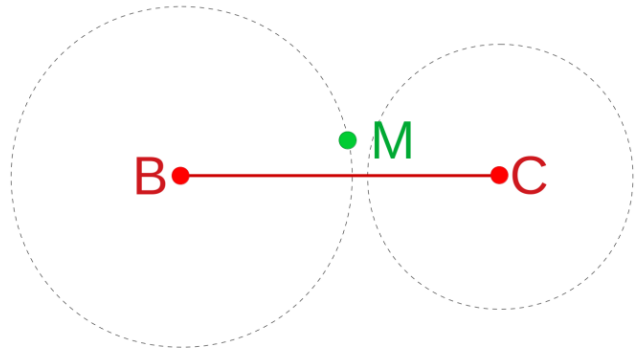

Figure 7. The underestimation difficulty.

Taking this behavior into account led to a new version of the algorithm: we start by making sure the estimated position estimation can only be located in the area between B and C, then if the sum of the estimated distances (B-M and C-M) is smaller than DBC, the algorithm reports the nodes as aligned. This resulted in a 100% success rate when the real angles are very small and did not create any false positives when the angles are greater than the threshold. The algorithm can therefore be summarized as: if the angles at $A_j$ and $A_k$ are below a given threshold and the locus of $X_i$ is in the bounding box between $A_j$ and $A_k$, $X_i$ is between $A_j$ and $A_k$.

It should be noted that some runs of the ranging algorithm return distance estimates which were clear outliers (over 1km): such samples were removed before processing. As a result, the number of samples per category varies.

### 5.3. Experiment Conclusion

Although the results were obtained using a testbed and without human interaction, they demonstrate the performance of the algorithms when it comes to the repeatability of the results.

## 6. Design pre-experiment in living lab

After an evaluation of the raw performance of the algorithms without considering the user's point of view, we report a second study to compare the return of the algorithms with the perception of objets localisation by real users.

### 6.1. Pre-experiment Design

We have set up a pre-experimentation protocol in the Maison Intelligente de Blagnac (MIB). The objective of this pre-experiment is to know how the user describes the position of a given object in relation to reference objects in three rooms (kitchen, living room and bedroom) of the MIB. We used two location modalities: 1) a location called proximity with respect to a reference object identified by the user; 2) a location between two reference objects (alignment).

#### 6.1.1. Population

Population: 10 able-bodied participants.

#### 6.1.2. Description of the situations

Four situations have been defined:
- Kitchen: two objects to locate: the smartphone (situation 1) and a bunch of keys (situation 2);
- Living room: two objects to locate: the glass (situation 3) and the TV remote control (situation 4);
- Bedroom: a pair of glasses (situation 5) is to be located.

Blue post-it notes are used to identify the reference objects. The reference objects are as follows:
- Kitchen: fridge, microwave oven, purple vase (on the table), white bowl (on the table), coffee maker (on the far left of the fixed countertop, against the wall), yellow kettle (on the left of the movable countertop, to the left of the drainer);
- Living room: television, fixed phone, robot vacuum cleaner, robot vacuum cleaner remote control, connected speaker (next to the sofa), wheelchair (between the sofa and the wall);
- Bedroom: remote control base, bed, bedside lamp, connected scale.

We defined different situations to correspond to more or less complex object localisation tasks depending on the reference object(s) selected. In order to compare the algorithm outputs to the participants' answers, we precisely measured using a laser rangefinder the distance between each object pair, regardless of their role (reference object or mobile object). An excerpt of the data is provided in Tab. 6 for the situation in figure 8: although the user is expected to only express the location of the remote control and glass, all distances have been stored. These distances have been measured in two ways:
- Edge to edge distance: the shortest distance between the outer perimeter (edge) of the two objects is stored,
- Inter-centre distance: the separation between the geometric centre of the objects is stored.

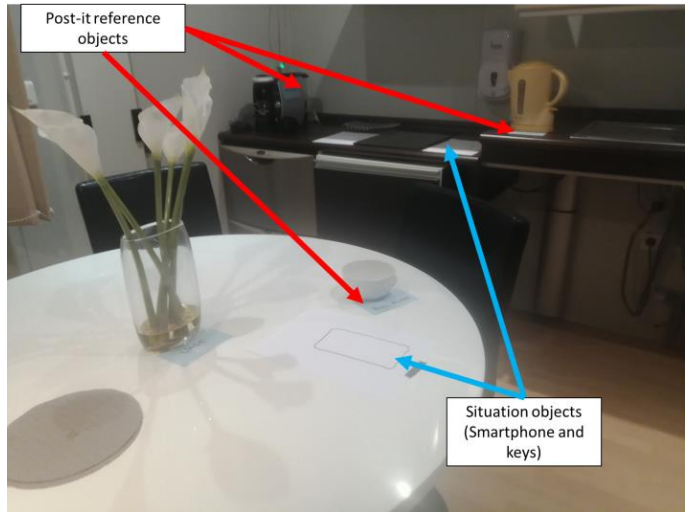

Figure 8. Objects in the kitchen.

Tab. 6. Example: edge to edge distances in the kitchen (cm).

|  | fridge | micro wave | vase | bowl | coffee maker | kettle | cell phone | keys |
|---|---|---|---|---|---|---|---|---|
| fridge | * |  |  |  |  |  |  |  |
| microwave | 50 | * |  |  |  |  |  |  |
| vase | 136 | 142 | * |  |  |  |  |  |
| bowl | 125 | 168 | 31 | * |  |  |  |  |
| coffee maker | 299 | 130 | 176 | 166 | * |  |  |  |
| kettle | 203 | 277 | 144 | 108 | 115 | * |  |  |
| cell phone | 105 | 147 | 23 | 11 | 185 | 130 | * |  |
| keys | 224 | 284 | 142 | 109 | 91 | 16 | 131 | * |

To ensure that all subjects have the same positioning in the room, we placed a black mark on the floor (as shown in Figure 9). Figure 8 gives an idea of situations 1 and 2 while figure 9 gives an idea of situations 3 and 4.

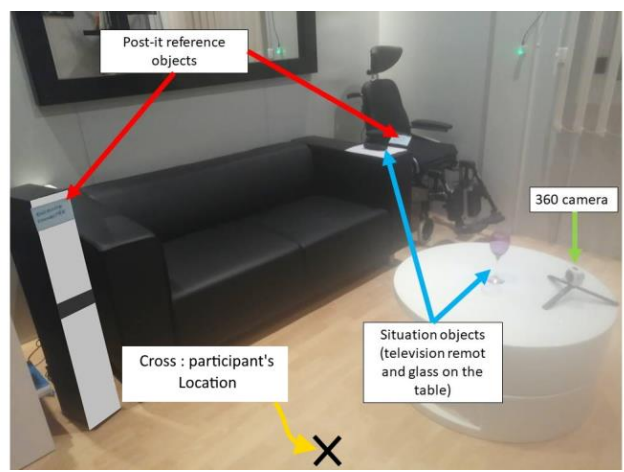

Figure 9. Objects in the livingroom.

At the beginning of each experiment, each participant had to sign a consent and image rights form. There was no request made to an ethics committee. Once this step was completed, the participant was given the following instructions: "For each situation, position yourself at the location indicated by the floor marker and then fill in the form to indicate the best description of the location of the target object in relation to other objects in the environment identified by a post-it. You may take a few seconds to prepare your answer".

*6.1.3. Methodological Tool*

The form consisted of the questions listed in Tab. 7.

Tab. 7. Questions.

| Question 1 | Question 2 |
|---|---|
| The object is located (only one choice)<br>❏ very close<br>❏ near<br>❏ in the vicinity of ...................................... (free choice of an object marked with a post-it note) | The object is located between ...............................<br>and .......................................<br><br>(free choice of two objects marked with a post-it note) |

The subject was free to answer as many times as he/she deemed relevant to locate the object in relation to the reference objects in each room. No instruction on completeness of location with respect to the totality of reference objects was given to the test subjects.

*6.2. Results*

In this study, we aimed to evaluate the relevance of our choices when implementing the proximity algorithm, namely the two rules (cf. § 5.1.2):
− R1: the 3 threshold values arbitrarily chosen (table III) at 0.3m (very close, VC) ; 0.6m (near, N) and 1.2m (in the vicinity, V). When the ranging is greater than 1.2m, the algorithm return an empty response (No Response -NR).
− R2: the fact of retaining only the smallest of the distances and eliminating the others at the output of the micro algorithm.

Tables (Tab. 8, Tab. 9, Tab. 10, Tab. 11 and Tab. 12) present a synthesis of the subjects responses to the question 1: "very close" (VC), "near" (N), "in the vicinity" (V) or "No response" (NR), for each situation.

The output of the proximity algorithm is also indicated in the tables, in the two considered cases "inter-centre" and "edge to edge distances".

The total number of possible responses for the location of the 5 objects is 280 for the five situations. We found 125 NR. The 155 answers correspond to the locations expressed by the 10 participants (Table VI). Some participants gave several locations per situation using one or more reference objects. When we consider only the "closest" answer we have only 49 answers. The analysis of the 155 responses shows that the participants developed different localisation strategies:
− 3 participants have given a description of the location of the smartphone in relation to the set of reference objects (situation 1) using adverbs expressing the various levels of physical distance; we observe the same strategy of these same participants for situation 2;
− For situation 3, 4 participants have also implemented this type of strategy but the adverb that is most used is around the TV and the wheelchair; we see that the answers are divided into 5 (N) answers against 4 (V) for the television and 7 (V) answers and 2 (N) answers for the wheelchair and one (NR);
− For situation 4, 90% of the answers correspond to (VC) against 10% to (N) of the wheelchair; 2 of the 10 participants have developed an exhaustive strategy of localisation of the remote control in relation to the reference objects;
− For situation 5, all 10 subjects responded "VC" with regards to the lamp; the two same participants as in situation 4 also developed an exhaustive strategy of localisation of glasses in relation to the other objects of reference.

Two participants implemented an exhaustive description of the locations regardless of the situation.

Now, if we analyze the 60 responses of situation 3, they are distributed between 23 (NR), 27 (V) and 10 (N). The responses in situation 1 are distributed between 14 (VC) answers, 8 (N), 7 (V) and 31 (NR). This distribution of responses for these two situations shows that participants seem to develop different strategies depending on the distance between objects.

In order to understand the participants' answers and to compare them with the Proximity Estimator algorithms' return, it is interesting to differentiate the two possibilities of interpretation of the users' "No Response" (NR): either (strategy 1) the users did not try to be exhaustive and stopped after having expressed n proximity relationships "by fatigue", or (strategy 2) they stopped at n proximity relationships because they considered that only these n expressions of proximities were relevant for their localisation strategy. The instructions given in our experiment do not allow us to identify the strategy used by the participants.

Tab. 8. Situation 1 results: cellphone.

| | **Smartphone** | fridge | microwave | vase | bowl | coffee maker | kettle |
|---|---|---|---|---|---|---|---|
| algo | inter-centre | NR | NR | N | VC | NR | NR |

|  | edge to edge distances | V | NR | VC | VC | NR | NR |
|---|---|---|---|---|---|---|---|
| subjects | VC | 0% | 0% | 40% | 100% | 0% | 0% |
|  | N | 10% | 10% | 50% | 0% | 0% | 10% |
|  | V | 20% | 10% | 10% | 0% | 20% | 20% |
|  | NR | 70% | 80% | 0% | 0% | 80% | 70% |

Tab. 9. Situation 2 results: keys.

| **Keys** |  |  | fridge | micro wave | vase | bowl | coffee maker | kettle |
|---|---|---|---|---|---|---|---|---|
| algo | inter-centre |  | NR | NR | NR | NR | V | N |
|  | edge to edge distances |  | NR | NR | NR | V | V | VC |
| subjects | VC |  | 0% | 0% | 0% | 0% | 0% | 90% |
|  | N |  | 0% | 0% | 0% | 20% | 70% | 10% |
|  | V |  | 20% | 20% | 30% | 30% | 30% | 0% |
|  | NR |  | 80% | 80% | 70% | 50% | 0% | 0% |

Tab. 10. Situation 3 results: glass.

| **Glass** |  |  | TV | fixed phone | RVC | RVC remote | speaker | wheel chair |
|---|---|---|---|---|---|---|---|---|
| algo | inter-centre |  | NR | NR | NR | NR | NR | NR |
|  | edge to edge distances |  | V | NR | NR | NR | NR | V |
| subjects | VC |  | 0% | 0% | 0% | 0% | 0% | 0% |
|  | N |  | 50% | 20% | 0% | 0% | 10% | 20% |
|  | V |  | 40% | 50% | 20% | 20% | 70% | 70% |
|  | NR |  | 10% | 30% | 80% | 80% | 20% | 10% |

Tab. 11. Situation 4 results: TV remote control.

| **TV remote control** |  |  | TV | fixed phone | RVC | RVC remote | speaker | wheel chair |
|---|---|---|---|---|---|---|---|---|
| algo | inter-centre |  | NR | NR | NR | NR | NR | N |
|  | edge to edge distances |  | NR | NR | NR | NR | NR | VC |
| subjects | VC |  | 0% | 0% | 0% | 0% | 0% | 90% |
|  | N |  | 0% | 0% | 0% | 0% | 40% | 10% |
|  | V |  | 20% | 40% | 20% | 20% | 40% | 0% |
|  | NR |  | 80% | 60% | 80% | 80% | 20% | 0% |

Tab. 12. Situation 5 results: glasses.

| **Glasses** |  |  | RC base | bed | bedside lamp | connected scale |
|---|---|---|---|---|---|---|
| algo | inter-centre |  | NR | NR | VC | V |
|  | edge to edge distances |  | NR | N | VC | N |
| subjects | VC |  | 0% | 30% | 100% | 20% |
|  | N |  | 0% | 50% | 0% | 30% |
|  | V |  | 20% | 10% | 0% | 20% |
|  | NR |  | 80% | 10% | 0% | 30% |

In the following paragraphs, we will analyse the influence of the second rule of the proximity algorithm: this parameter also impacts the completeness of the answers as only the smallest distance is kept under rule n°2.

When considering user strategy 1, the NRs of the participants are voluntarily excluded: only the responses VC, N and V are considered. For the 10 subjects and the 5 objects to be located, the subjects expressed the same feedback as the proximity algorithm in 20.6% of the situations (32 out of 155 expressed responses) for distances between objects from geometric center to geometric center. This value rises to 44.5% (69 out of 155 expressed responses) if we consider distances between objects from edge to edge.

Still excluding NRs, if we also consider rule 2 of the Proximity Estimator algorithm, the 10 subjects expressed the same feedback as the algorithm in 44.9% of cases (22 out of 49 expressed responses) if we consider distances between objects from geometric center to geometric center. This value rises to 92.8% (45 out of 49 expressed responses) if we consider distances between objects from edge to edge.

Now, if we consider the participants' NRs as the algorithm does, i.e. that beyond 1.2m, a reference object is considered too far away to give a relevant indication, the results converge 55% (154 out of 280 possible answers) of the time if we consider the distances between objects from geometric center to geometric center; this value rises to 62.9% (176 out of 280 possible answers) if we consider the distances between objects from edge to edge.

Finally, if we also consider rule 2, participants expressed the same return as the algorithm 46% of the time (23 out of 50 possible answers) if we consider distances between objects from geometric center to geometric center; this value is 90% (45 out of 50 possible answers) if we consider distances between objects from edge to edge.

*6.3. Discussion*

In this section, we discuss the results from both automated testing and experiments involving human interaction.

The testbed-based evaluation of the algorithms highlighted the impact of the chosen technology on performance. Although ToF-based UWB ranging is one of the most precise and readily available solutions, in close proximity, the performance drops. In order to fix the issue, two methods can be used: the technology-centric method involves implementing a ranging correction model that will focus on the ranging error when, the distance is below 1.5m. The second method takes into consideration the human point of view: taking the ranging tool as is, experiments with real users selected from the target population will be run with a hidden contributor pulling the strings. This wizard of Oz will be observing the scene via cameras and will select in real-time which DSP to use when informing the participant. At the end of the experiment, the degree of satisfaction of the participant will be collected in order to adjust the thresholds.

Regarding the experiments involving human participants, the results clearly favour the edge-to-edge distance calculation. Entering the size of the object in the proximity estimation clearly increased the relevance of the expressions generated by the algorithm.

The results also show that if the users' perception of "VC" is quite close to the threshold arbitrarily chosen for the algorithm, an increase of the two other thresholds would be beneficial: the higher bound of "V", in particular, is perceived as being much further than 1.2m in a significant number of situations.

The threshold refinement process must also take into consideration the free space between objects.

The analysis of object localisation by participants also shows that participants develop different localisation strategies: some attempt to exhaustively describe the situation while others favor localisation by more discriminating adverbs ("VC" or "N").

These two strategies suggest that the implementation of the proximity algorithm should rely on a set of objects relevant to the disabled person according to his/her abilities. For example, for a person with cognitive disorders, the final information provided may be restricted to reference objects located in the same room.

Of course, we are aware that these initial results were obtained as part of a pre-experiment carried out with participants without disabilities. These first results presented here will have to be validated later. We will carry out a larger-scale evaluation on participants with visual and/or cognitive disabilities in order to be able to confirm/rebut these initial results and, possibly, to enrich the field of knowledge.

**Conclusion and perspectives**

In this article, we have applied localisation techniques to the field of health and autonomy. We have illustrated the demands of the disabled community through two concrete cases of visual impairment and cognitive impairment. Several usecases of object localisation have been identified. An overview of localisation technologies using connected objects was presented; beyond the technologies, fundamental concepts allowing to differentiate the technologies were presented: range, coordinate system, architecture and localisation outputs. We then introduced the concept of semantic localisation built on top of ranging service based on UWB. The solution consists in a wireless network of smart devices which generate distance estimates that are fed to micro-algorithms. Those micro-algorithms produce pieces of semantic localisation information that are then combined by a fusion algorithm. We described Room Determination, Proximity Estimator and Alignment Estimator algorithms: Proximity Estimator and Alignment Estimator algorithms have been evaluated using a testbed and the results have been compared to those obtained from a measurement campain involving 10 able-bodied human participants. The results are encouraging and, in some cases, both experiments converge up to 90% of the time.

A first perspective of this work is the full comparison of the three proposed algorithm. The deployment of the LocURa4IoT testbed in the Maison Intelligente de Blagnac makes this full comparison possible.

A new user experiment is underway to identify the location adverbs used. This experiment will provide insights on participants' localisation strategies and will suggest other potential localisation micro-algorithms (e.g., next to, in front of, etc.). Therefore, new algorithms are being investigated in order to introduce other types of SPD.

Some of these future algorithms require the ability to activate an ego-centric localisation mechanism which is only possible if the user's location and orientation are known to the system. This exciting challenge is one of the next steps of the IDEALI project and will benefit from the techniques based on Angle of Arrival (AoA).

Although it is not the focus of the work presented here, we are aware that we will also need to study the effects of depth perception on the ability of sighted subjects to assess distances between themselves and reference objects in the scene. Binocular convergence and visual accommodation work for distances less than 20 meters but are fairly inaccurate visual cues. The familiar size of objects, the effects of interposition, the perspectives constitute a reference frame acquired from childhood. The parallax of head movements is a major clue that informs us about the relative situation of objects. Finally, an environment constituted by heterogeneous surfaces with gradients of colors and textures provides important information concerning the perception of distance. In a future study, we will vary certain visual cues and observe the impact of these variations on the subjects' ability to evaluate the distance between themselves and the different reference objects.

These perspectives require a switch to a transdisciplinary approach in the future. Finally, the location of connected objects extends the field of possibilities: new innovative applications are possible, making it possible to improve the daily life of users, with or without disabilities.

**Acknowledgement**

This study was supported by the CNRS INS2I department in 2019.

**References**

[1] S. Katz, A.B. Ford, R.W. Moskowitz, B.A. Jackson, M.W. Jaffe, "Studies of the illness in the aged. The index of ADL: a standardized measure of biological", JAMA 1963; 21: 914-9.

[2] M.P. Lawton, E.M. Brody, "Assessment of older people: Self-maintaining and instrumental activities of daily living", Gerontologist 1969. Doi:10.1093/geront/9.3_Part_1.179.

[3] A. van den Bossche, N. Gonzalez, T. Val, D. Brulin, F. Vella, N. Vigouroux & E. Campo, Specifying an MQTT Tree for a Connected Smart Home. Mokhtari M.; Abdulrazak B.; Aloulou H. (eds). Smart Homes and Health Telematics, Designing a Better Future: Urban Assisted Living. ICOST 2018. Lecture Notes in Computer Science, vol 10898., Springer, pp.236-246, 2018, ⟨10.1007/978-3-319-94523-1_21⟩. ⟨hal-01997243⟩

[4] F. Dramas, S. Thorpe, C. Jouffrais. "Localisation d'objets pour les non-voyants par analyse d'image : analyse du besoin et prototypage », Conférence Internationale sur l'accessibilité et les systèmes de suppléance aux personnes en situations de handicaps (ASSISTH 2007), Toulouse, France, Cépaduès, pp. 109-116, novembre 2007

[5] L. Hamilton, S. Fay, K. Rockwood. "Misplacing objects in mild to moderate Alzheimer's disease: a descriptive analysis from the VISTA clinical trial", J Neurol Neurosurg Psychiatry, vol. 80, no.9 p.960—5, 2009.

[6] B. Boudet, T. Giacobinia., I. Ferrané, C. Fortin, C. Mollaretb, F. Lerasle, P. Rumeau, « Quels sont les objets égarés à domicile par les personnes âgées fragiles ? Une étude pilote sur 60 personnes », NPG Neurologie - Psychiatrie - Gériatrie vol.14, pp.38—42, 2014.

[7] http://www.alzheimer-europe.org/EN/Living-with-dementia/ Caring-for-someone-with-dementia/Changes-in-behaviour/Hiding-losing-objects-and-making-false-accusations#fragment

[8] A. Shelton, T. MacNamara, "Systems of Spatial Reference in Human Memory", Cognitive Psychology, vol. 43, pp.274-310, 2001

[9] R. Klatzky, "Allocentric and egocentric spatial representations: Definitions, distinctions, and interconnections", In C. Freksa, & C. Habel, (Eds.) Wender, Spatial cognition - An interdisciplinary approach to representation and processing of spatial knowledge, Berlin: Springer- verlag. pp. 1-17, 1998

[10] W. Mou, T. McNamara, « Intrinsic Frames of Reference in Spatial Memory", Journal of experimental psychology: Learning, Memory and cognition, vol. 28, pp.162-170, 2002

[11] S. Feigenbaum, R. Morris, "Allocentric versus egocentric spatial memory after unilateral temporal lobectomy in humans", Neuropsychology, vol. 18, pp.462-472, 2004

[12] E. Coluccia, I. Mammarella, C. Cornoldi, "Centred egocentric, decentred egocentric, and allocentric spatial representations in the peripersonal space of congenital total blindness", Perception, vol. 38, pp.679-693, 2009

[13] M. Simonnet, S. Vieilledent, J. Tisseau, « Note théorique : Influences des activités du sujet et des caractéristiques environnementales sur la nature de l'encodage spatial », Année Psychologique, Centre Henri Pieron/Armand Colin, vol. 113 no.2, pp.227 – 254, 2013

[14] A. Brilhault, « Vision artificielle pour les non-voyants : une approche bio-inspirée pour la reconnaissance de formes ». Intelligence artificielle [cs.AI], Université Toulouse III Paul Sabatier, Français. tel-01127709, 2014

[15] P. Rumeau, B. Boudet, C. Mollaret, I. Ferrané, F. Lerasle, « Etude de l'IHR sur deux groupes de personnes agées », 28ième conférence francophone sur l'Interaction Homme-Machine, Fribourg, Suisse, pp.13-15, Oct 2016

[16] P. Lopes, M. Pino, G. Carletti, S. Hamidi, S. Legué, H. Kerhervé, S. Benveniste, G. AndÃ©ol, P. Bonsom, S. Reingewirtz, A.-S. Rigaud, "Co-Conception Process of an Innovative Assistive Device to Track and Find Misplaced Everyday Objects for Older Adults with Cognitive Impairment: The TROUVE Project", IRBM, vol.3, no.2, pp.52-57, 2016, ISSN 1959-0318,https://doi.org/10.1016/j.irbm.2016.02.004.

[17] K. Müller, S. Das, "REANA : An RFID-Enabled Environment-Aware Navigation System for the Visually Impaired", 2010

[18] L.A. Guerrero, F. Vasquez, S.F. Ochoa, "An Indoor Navigation System for the Visually Impaired" Sensors, vol. 12, no.6, pp.8236-8258, 2012 https://doi.org/10.3390/s120608236

[19] F. Dramas, S. Thorpe, C. Jouffrais, "Artificial vision for the blind: a bio-inspired algorithm for objects and obstacles detection". International Journal of Image and Graphics, vol. 10, no.04, pp. 531-544, 2010

[20] M. Macé, G. Denis, C. Jouffrais, "Simulated prosthetic vision: The benefits from computer based object recognition and localisation" Artificial Organs, vol. 39, no.7, pp. E102-E113, 2015 ISSN 0160-564X

[21] F. Dramas, B. Oriola, B. Katz, S. Thorpe, C. Jouffrais, "Designing an assistive device for the blind based on object localisation and augmented auditory reality", the 10th international ACM SIGACCESS conference, Halifax, France, Oct 2008

[22] D. Appert, D. Camors, J.B. Durand, C. Jouffrais, « Assistance tactile à la localisation de cibles périphériques pour des personnes à vision tubulaire », IHM 2015 27ème Conférence Francophone sur l'interaction Homme-Machine, Toulouse, France, pp.1-10, Oct 2015

[23] Brock, S. Kammoun, M. Macé, C. Jouffrais, "Using wrist vibrations to guide hand movement and whole body navigation", i-com Zeitschrift für interaktive und kooperative Medien, vol. 13, no.3, pp.19-28, 2014 ISSN 1618-162X

[24] M. Asma Ben Hadj, T. Val, L. Andrieux, A. Kachouri, « A Help for Assisting People based on a Depth Cameras System Dedicated to Elderly and Dependent People », Journal of Biomedical Engineering and Medical Imaging, Scholar Publishing, vol. 1, no.6, December 2014

[25] https://www.challenge-malin.fr/

[26] F. Bettahar, W. Bourennane, Y. Charlon, E. Campo, « HOMECARE: une plateforme technique de surveillance pour le suivi actimétrique de patients Alzheimer », Workshop – Alzheimer, Approche pluridisciplinaire, De la recherche clinique aux avancées technologiques, Jan 2013

[27] A. Marco, R. Casas, J. Falco, H. Gracia, J.I. Artigas, A. Roy, « Location-based services for elderly and disabled people », Elsevier Comuter communications, Janv 2008

[28] R. Dalce, "Méthodes de localisation par le signal de communication dans les réseaux de capteurs sans fil en intérieur", thèse de doctorat, université de Toulouse, 2013

[29] P. Barsocchi, F. Furfari, P. Nepa, « RSSI localisation with sensors placed on the user », Indoor Positioning and Indoor Navigation (IPIN), october 2010

[30] F. Despaux, A. Van Den Bossche, K. Jaffres-Runser, T. VAL, "N-TWR: An Accurate Time-of-flight-based N-ary Ranging Protocol for Ultra-Wide Band", Ad Hoc Networks Journal, Elsevier, juin 2018

[31] Thèse Géolocalisation d'émetteurs en une étape : Algorithmes et performances, Cyrile Delestre, fev 2016, https://tel.archives-ouvertes.fr/tel-01280408/document

[32] plateforme Bluetooth 5.1, Digi-Key. https://www.digikey.fr/fr/articles/use-bluetooth-5-1-enabled-platforms-part-1

[33] composants radio UWB AoA. https://www.qorvo.com

[34] D. Michel, « Descriptions spatiales », Petit traité de l'espace. Un parcours pluridisciplinaire, sous la direction de Denis Michel. Wavre, Mardaga, « PSY-Théories, débats, synthèses », pp.173-198, 2016  URL : https://www.cairn.info/petit-traite-de-l-espace--9782804703226-page-173.htm

[35] A. Van Den Bossche, R. Dalce, N. Gonzales, T. Val, "LocURa: A New Localisation and UWB-Based Ranging Testbed for the Internet of Things", IEEE International Conference on Indoor Positioning and Indoor Navigation (IPIN 2018), Nantes, France, sept 2018

[36] http://mi.iut-blagnac.fr/?lang=fr

[37] https://www.irit.fr/~Adrien.Van-Den-Bossche/media/ideali1-demo.mp4

[38] A. van Den Bossche., R. Dalcé, T. Val, "LocURa4IoT -A testbed dedicated to accurate localisation of wireless nodes in the IoT", IEEE Sensors Journal, Institute of Electrical and Electronics Engineers, In press ,Accès: https://hal-univ-tlse3.archives-ouvertes.fr/hal-03466307